\documentstyle[12pt,aasms4]{article}
\slugcomment{\em To appear in Astrophysical Journal, Letters to the 
    Editor}
\begin{document}

\title{Substructure in the Globular Cluster System of the Milky Way:
       The Highest-Metallicity Clusters}

\author{Andreas Burkert}
\affil{Max-Planck-Institut fur Astronomie, K\"onigstuhl 17,\\
       D-69117 Heidelberg, \\ Germany }
\author{Graeme H. Smith}
\affil{University of California Observatories/Lick Observatory, \\ 
       University of California,\\ Santa Cruz, California 95064}
\authoremail{burkert@mpia-hd.mpg.de}

\begin{abstract}
An analysis of the kinematical and spatial properties of the 
highest-metallicity globular clusters in the Galaxy, having metallicities of 
[Fe/H] $> -0.8$, indicates that these objects do not comprise a homogeneous
population. Three subsystems are identified among these clusters.\\
{\it i)} The highest-mass clusters with $\log (M/M_{\sun}) > 5.5$ exhibit a 
   very slow net rotation with a speed of $v_{\rm rot} = 24 \pm 23$ km/s and
   $v_{\rm rot} / \sigma_{\rm los} = 0.3$, indicative of a
   a centrally condensed, relatively-high-metallicity subsystem.\\
{\it ii)} Roughly half of the lower-mass clusters 
   appear to be located in an elongated bar-like structure which passes 
   through the Galactic Center, and has similar properties to the
   central stellar bar of the Milky Way.\\
{\it iii)} The remaining lower-mass clusters exhibit very rapid net rotation, 
    with a rotation speed of $v_{\rm rot} = 164 \pm 6$ km/s and 
    $v_{\rm rot} / \sigma_{\rm los} = 6$. These clusters are
    located in the Galactic plane, within a ring of 4 to 6 kpc radial distance
    from the Galactic Center.

The highest-mass clusters may have formed during relatively advanced stages of 
the dissipative evolution of the inner Galactic halo. Although the lower-mass 
bar clusters have kinematical properties which are similar to
the highest-mass clusters, their spatial distribution
suggests that they may be associated with the formation of the 
Galactic stellar bar or bulge. The lower-mass ring 
clusters appear to be real disk 
objects. They may represent a stage in cluster formation that was intermediate 
between that of the halo globular clusters and the oldest extant open clusters.
\end{abstract}


\section{Introduction}
Since the pioneering work of Shapley (1918) the properties of the Galactic 
globular cluster system have been seen as valuable tracers of the structure and
history of the Galactic halo (van den Bergh 1995). 
Zinn (1985), Armandroff \& Zinn (1988),
and Armandroff (1989), on the basis of correlations between metallicity
and kinematics, divided the globular cluster population 
of the Galaxy into a halo and a disk system, the 
clusters of which have metallicities of [Fe/H] $< -0.8$ and [Fe/H] $> -0.8$,
respectively. This has lead to a model in which the metal-richest globular 
clusters belong to a kinematically uniform disk system.
This letter presents a new analysis of the kinematical and 
spatial properties of the highest-metallicity ([Fe/H] $> -0.8$) Galactic 
globular clusters which shows that they may not 
constitute a homogeneous population. Instead, if these clusters are grouped on
the basis of their mass, several subsystems can be identified.
We have taken advantage of the April 1996 version of the {\it Catalog of 
Parameters for Milky Way Globular Clusters} compiled by Harris (1996). 
This was obtained through the World Wide Web from the site (URL) 
http://www.physics.mcmaster.ca/Globular.html.
Our kinematical analysis follows the 
lines of Frenk \& White (1980, 1982), Zinn (1985), and
Armandroff (1989), but our approach differs from 
theirs in that we divide the metal-richest clusters into two subgroups 
according to cluster mass.
  
\section{Mathematical Framework}

Throughout this paper a standard $X, Y, Z$ Galactocentric coordinate system is
adopted (cf. Mihalas \& Binney 1981).
The Solar motion is taken to be in the direction $l_s =$ 57\arcdeg, 
$b_s =$ 23\arcdeg, with a velocity $S = 20$ km/s (Zinn 1985).
The three components of the Solar velocity relative to the 
Local Standard of Rest (LSR) are
then $(U_{\sun}, V_{\sun}, W_{\sun}) = (10.1, 15.5, 7.5)$ km/s.
The velocity of a globular cluster with respect to the LSR 
($V_{\rm LSR}$) is computed from the above components of the Solar motion and
the heliocentric radial velocity of the cluster (Mihalas \& Binney 
1981).  The radial velocity $V_{\rm S}$ of a cluster with respect to
a {\it stationary} observer at the location of the Sun is
$V_{\rm S}  =  V_{\rm LSR} + 220 \phantom{i}{\rm (km/s)} \cos A,$
where $A$ is the angle between the apex of the LSR and the cluster's 
position on the sky, $\cos A  =  \sin l \cos b$ 
and the rotational velocity of the LSR is assumed to be 220 km/s.

Analysis of the bulk rotation of a system of globular clusters
depends on that system's rotation curve.
In the case of a constant systematic rotation speed $v_{\rm rot}$ 
without a global radial motion (Frenk \& White 1980)

\begin{equation}
            V_{\rm S} = v_{\rm rot} \cos \Psi + v_{\rm pec}
\end{equation}

\noindent where 

\begin{equation}
\cos \Psi = \frac{R_{\sun} \cos A}{\sqrt{R^2 \cos^2 A + (R_{\sun}-R 
            \cos b \cos l)^2}},
\end{equation}

\noindent $v_{\rm pec}$ is the line-of-sight peculiar velocity of a 
cluster, and $R$ is the cluster's Galactocentric distance.
The Galactocentric distance of the Sun is taken to be $R_{\sun} = 8$ kpc. 
In the case of rigid-body rotation with a uniform angular velocity of
$\omega$ (Kinman 1959, Zinn 1985)

\begin{equation}
              V_{\rm S}  =  \omega  R_{\sun}  \cos A + v_{\rm pec}.
\end{equation}

\noindent An unbiased estimate of $v_{\rm rot}$ can be obtained by the 
following sum over all clusters (Frenk \& White 1980):

\begin{equation}
v_{\rm rot} = \frac{ \sum_i \cos \Psi_i \ V_{\rm S,i}}{\sum_i \cos^2 \Psi_i}
 \pm \frac{\sigma_{\rm los}}{\sqrt{\sum_i \cos^2 \Psi_i}}
\end{equation}

\noindent with a similar equation for $\omega R_{\sun}$ and $\cos A$.
The line-of-sight velocity dispersion $\sigma_{\rm los}$ of the sample is
determined by inserting $v_{\rm rot}$ or $\omega R_{\sun}$ into 
equations 1 and 3, respectively, and calculating the dispersion in
$v_{\rm pec}$.
 
Globular cluster masses $M$ were calculated from the integrated absolute visual 
magnitudes ($M_{\rm V}$) tabulated by Harris (1996). A mass-to-light ratio of 
$(M/L)_{\rm V} = 3$ was assumed (Chernoff \& Djorgovski 1989).

\section{The Two Subsystems Among the Highest-Metallicity Globular Clusters}

In this section we divide the metal-richest Galactic globular clusters, which 
we take to be those with [Fe/H] $> -0.8$, 
into two mass groups: a {\it high-mass} group for which 
$\log M/M_{\sun} > 5.55$ ($M_{\rm V} < -7.9$), 
and a {\it lower-mass} group, containing clusters of 
mass smaller than this. The kinematical properties of these two groups are 
summarised in columns 2 and 3 of Table~1. The parameters listed include the 
number of clusters in each group $N$, the
rotation speed $v_{\rm rot}$ for the case of a flat rotation curve and
the value of $\omega $ for the case of rigid-body rotation, as well
as the standard deviations in these values.
Also given are the line-of-sight velocity dispersions $\sigma_{\rm los}$,
the mean height above the Galactic plane $Z_{\rm mean}$, and the 
mean radial distance in the Galactic plane $R_{\rm mean}$ for each group.
With the caveat that the number of clusters is 
relatively modest, the entries in Table~1 indicate that the kinematical
properties of the high-mass and lower-mass clusters are different.
While the high-mass clusters show little or no rotation, the low-mass
clusters exhibit rotation with a significance level of
$v_{\rm rot} \approx 8 \sigma$. These results suggest that the
high-mass clusters define a spheroidal inner-halo-like subsystem
whereas the low-mass clusters represent a disk-like component.

\subsection{The Lower-Mass Clusters: $\log M/M_{\sun} < 5.55$}

Kinematical and spatial data for the lower-mass clusters are plotted in the 
upper and lower panels, respectively, of Figure~1. Least-squares fits to the 
velocity data for the entire set of lower-mass clusters are shown as 
short-dashed lines in the upper panels.
Two groups of clusters can be distinguished in the top panels of Figure~1. One
group (open circles) exhibits a relatively well-defined relationship between 
$V_{\rm S}$ and both $\cos A$ and $\cos \Psi$. The remaining clusters are all
within a small $\cos A$ range ($-0.2 < \cos A < 0.2$) and are represented by 
filled circles. These two groups are also relatively distinct in the lower 
panels of Figure~1. With the exception of Ter 7 which is located at 
Galactocentric coordinates $(X,Y,Z) = (-12.5,1.2,-7.5)$ kpc and probably 
belongs to the Sagittarius dwarf spheroidal galaxy (Da Costa \& Armandroff 
1995), the group with small $\cos A$ falls within a 
centrally concentrated, bar-like configuration
that is extended in the $X$ direction and has $Y \approx 0$; the clusters in
this group will be referred to as ``bar clusters.''
The remaining clusters (open circles) fall within a ring around the Galactic 
Center having an inner and an outer radius of $\approx$ 4 and 6 kpc 
respectively; these objects will be referred to as the ``5-kpc-ring'' clusters.

A bar-like feature in the space distribution of globular clusters was noted by
Woltjer (1975) and Harris (1976), who suggested that it could be an artifact of
distance measurement errors for clusters in directions close to that of the 
Galactic Center. Frenk \& White (1980) performed
Monte Carlo simulations to assess this possibility. They 
found that errors in distance moduli would have to be as large as 1.0 mag to
explain the bar. Whereas the accuracy of distance modulus measurements have 
increased since the time of Woltjer's (1975) analysis, the
bar is more prominent in Figure~1 than in the data of Woltjer, the opposite to
what would be expected if the bar were an observational artifact.
Note that there is also strong evidence for a
bar-like distribution  of stars and molecular gas in the Galactic
bulge (see the review by Freeman 1996). It is striking that the stellar bar
has a very similar orientation, spatial distribution, and 
rotational velocity to the subsystem of ``bar clusters'', indicating
a common origin.

Separate least-squares fits were made to the velocity data for the subset of
``bar'' clusters and for the subset of ``5-kpc-ring'' clusters. The results are
given in columns 4 and 5 of Table~1, and the fits for the ``5-kpc-ring'' 
subsystem are shown as long-dashed lines in Figure~1. Two points deserve 
emphasis.

{\it (i)} The small range of $\cos A$ for the {\it bar clusters}
leads to large errors in $\omega $ if one assumes solid-body rotation.
Nevertheless, the value of $\omega = 59$ km/s/kpc for these clusters
is in excellent agreement with Zhao's (1994) estimated angular velocity 
of the central stellar bar of $\omega = 60$ km/s/kpc.
The assumption of rigid body rotation might therefore be the
physically appropriate model for the ``bar clusters.''
It is interesting that all the clusters of the bar system
have positive $V_{\rm S}$ values.  This bulk flow
may indicate that the bar is a transient phenomenon.

{\it (ii)} The {\it 5-kpc-ring} clusters constitute a rapidly rotating 
subsystem, with only a small velocity dispersion. These clusters seem to be
members of the Galactic disk. The rotation speed at $R=5$ kpc, derived
from the solid-body assumption (upper-right panel of Figure 1) is 120 km/s, 
which compares with 164 km/s under the assumption of a flat rotation curve.
In addition, $v_{\rm rot}/\sigma_{\rm los} = 6$, indicative of a disk with 
well-organised rotation. This subgroup of ``5-kpc-ring'' clusters
has kinematics not greatly different from the old open cluster system of the 
Galactic disk, which has a scale height of 375 pc (Janes \& Phelps 1994) and
kinematics of $v_{\rm rot} = 211 \pm 7$ km/s and $\sigma_{\rm los} = 28$ km/s
(Scott, Friel, \& Janes 1995). The difference in the rotation speeds of these 
two cluster systems may be partly related to the differences in their average 
age and Galactocentric distance.

The lower-mass ``5-kpc-ring'' globular clusters might
represent ``missing links'' between the globular clusters of the
halo and the open clusters of the Galactic disk. As illustrated in Figure~3
of Hufnagel \& Smith (1994), for example, no old open clusters are known with
Galactocentric distances of less than 7~kpc. It may be that old
open clusters did not form within 7~kpc of the Galactic Center; and that more
populous globular-like clusters, like M71, were instead 
formed in this region of the Galaxy. Alternatively, open 
clusters that formed at $R < 7$~kpc may have been disrupted
by interactions with molecular clouds. The objects more likely to 
survive this process would have been the higher-mass globular clusters.

\subsection{The High-Mass Clusters: $\log M/M_{\sun} > 5.55$}

The positional and kinematical data for those globular clusters 
having $\log M/M_{\sun} > 5.55$ and [Fe/H] $> -0.8$ are shown in Figure~2.
The fact that $v_{\rm rot}/\sigma \leq 1$ indicates that the high-mass, 
metal-richest globulars are not members of a true rotating disk system.
The formal solution of $v_{\rm rot} = 24.3$ km/s is heavily weighted by the
cluster NGC~104 (47~Tuc) with $\cos \Psi = -0.68$ and $V_{\rm S} = -145$ km/s; 
excluding this object would lead to counter-rotation with $v_{\rm rot} = -10$ 
km/s for the remaining sample of clusters, further strengthening
the conclusion that these clusters are not part of a
rotating disk population.

Rather than being members of the disk, these clusters may 
instead belong to a metal-enriched inner-halo component. 
The velocity dispersion among these clusters is not as large as for the system
of more metal-poor Galactic halo globular clusters. 
This result is in agreement with
Frenk \& White (1980) who found that the observed line-of-sight velocity
dispersion of the halo cluster population increases significantly with
Galactocentric distance, indicating a non-isothermal cluster distribution
function. The high-mass 
relatively-high-metallicity clusters might therefore be part of a more 
centrally condensed and kinematically cooler inner component of the Galactic 
halo. Formed at times when the interstellar medium in the proto-galaxy had 
experienced dissipational contraction and
the metallicity had increased to relatively high values, these clusters may
represent the last remnants from the true epoch of globular cluster formation
in the Galactic halo. 

\subsection{Statistical Analysis}

A statistical investigation of the difference between the low-mass and 
high-mass clusters depicted in Figures~1 and 2, respectively, is affected
by the low number (7) of high-mass clusters. To take account of
this limitation, Monte Carlo simulations were performed in which 
random samples of seven clusters were generated 
from a population with given average rotational velocity
$V_{pop}$ and velocity dispersion $\sigma_{pop}$
similar to that of the low-mass clusters.
For each sample generated, the average rotational velocity $v_{\rm rot}$ 
was calculated using equation 4. The results for a large number of artificial 
samples were combined to give the probability that the rotation speed 
for such a sample is $v_{\rm rot} < 24.3$ km/s, corresponding to
rotation that is as slow as observed for the high-mass 
system of clusters. The data for each set 
of seven clusters were generated as follows: first, a value
for $\cos \Psi$ was 
chosen, randomly distributed between $-1$ and +1. The cluster velocities 
$V_{\rm S}$ were then generated assuming a normal probability 
distribution function:

\begin{equation}
      P(V_{\rm S}) = exp \left( - \frac{[V_{\rm S} - V_{pop} \cos \Psi ]^2}
               {2 \ \sigma_{pop}^2 } \right)
\end{equation}

\noindent with values of $V_{pop}$ and $\sigma_{pop}$ chosen to be 
representative of the low-mass system of clusters.

Using this method, we find that there exists a 10\% probability that the 
high-mass clusters are drawn from the same kinematical population as the 
combined sample of low-mass clusters.
Comparing the high-mass clusters and the low-mass bar clusters, we find
a 51\% probability that these two samples have similar
kinematical properties. On the other hand, the velocity distributions of 
high-mass clusters and low-mass ring clusters are significantly
different, with a probability of less than $4 \times 10^{-6}$ that both samples
result from the same population.

In summary, we can rule out with a high probability that the high-mass
clusters have an average rotational velocity and a velocity 
dispersion similar to the ring clusters. They appear to be two distinct
subsystems. The relation of the bar clusters to the high-mass clusters is not 
clear from the present analysis. The similar kinematical properties of both
subsystems (assuming a constant rotational velocity) suggests that they may be 
drawn from the same or related populations. However, the more extended spatial 
distribution of the high-mass clusters indicates that this subsystem 
is part of the inner halo. The origin of the bar clusters, on the 
other hand, is more likely related to the formation of the central stellar bar 
or bulge. The formation of the inner halo and Galactic bulge/bar 
could also be coupled. 

\section{Summary}

We propose that the following three subsystems can be distinguished among the
highest-metallicity ([Fe/H] $> -0.8$) globular clusters of the Galaxy:
{\it i) a centrally condensed halo subsystem of high-mass clusters,
ii) a bar subsystem of lower-mass clusters, and
iii) a rapidly-rotating disk subsystem of lower-mass clusters, many of
which fall within a ring about the Galactic Center of radius 4-6 kpc.}
These subsystems indicate that the population of relatively-high-metallicity 
globular clusters is more heterogeneous than has hitherto been expected from 
the concept of a single disk population. 

If the high-mass, higher-metallicity globular clusters really are 
members of the Galactic halo as suggested above, then it follows that 
high-mass ($M >3 \times 10^5$ $M_{\sun}$) cluster formation
only took place in the Galactic halo,
not in the disk. This suggests that the maximum mass of star clusters which 
formed in the Galaxy may have decreased with time, perhaps along the lines of 
the following scenario. The halo supported massive cluster formation, possibly 
under starburst-like conditions in massive clouds (Murray \& Lin 1989;
Harris \& Pudritz 1994; Brown et~al. 1995; Burkert et~al. 1996).
However, as the interstellar medium collapsed into the center and the
equatorial plane and the metallicity built up, cluster formation tended 
to produce the low or intermediate-mass globular clusters 
($\log M/M_{\sun} < 5.5$) found in the bar or the 4-6 kpc ring noted in 
Figure~1. As the disk built up further from the inside out,
the sites of cluster formation propagated outwards through the
disk into regions beyond 7~kpc from the Galactic Center. The typical masses of 
clusters being formed at this time were markedly lower, with open clusters
like NGC~6791 and NGC~188 being produced. 

\acknowledgments

A.~Burkert thanks the staff of UCO/Lick Observatory for their hospitality 
during two visits on which this work was done.

\newpage
\centerline{\bf Figure Captions} 

\figcaption[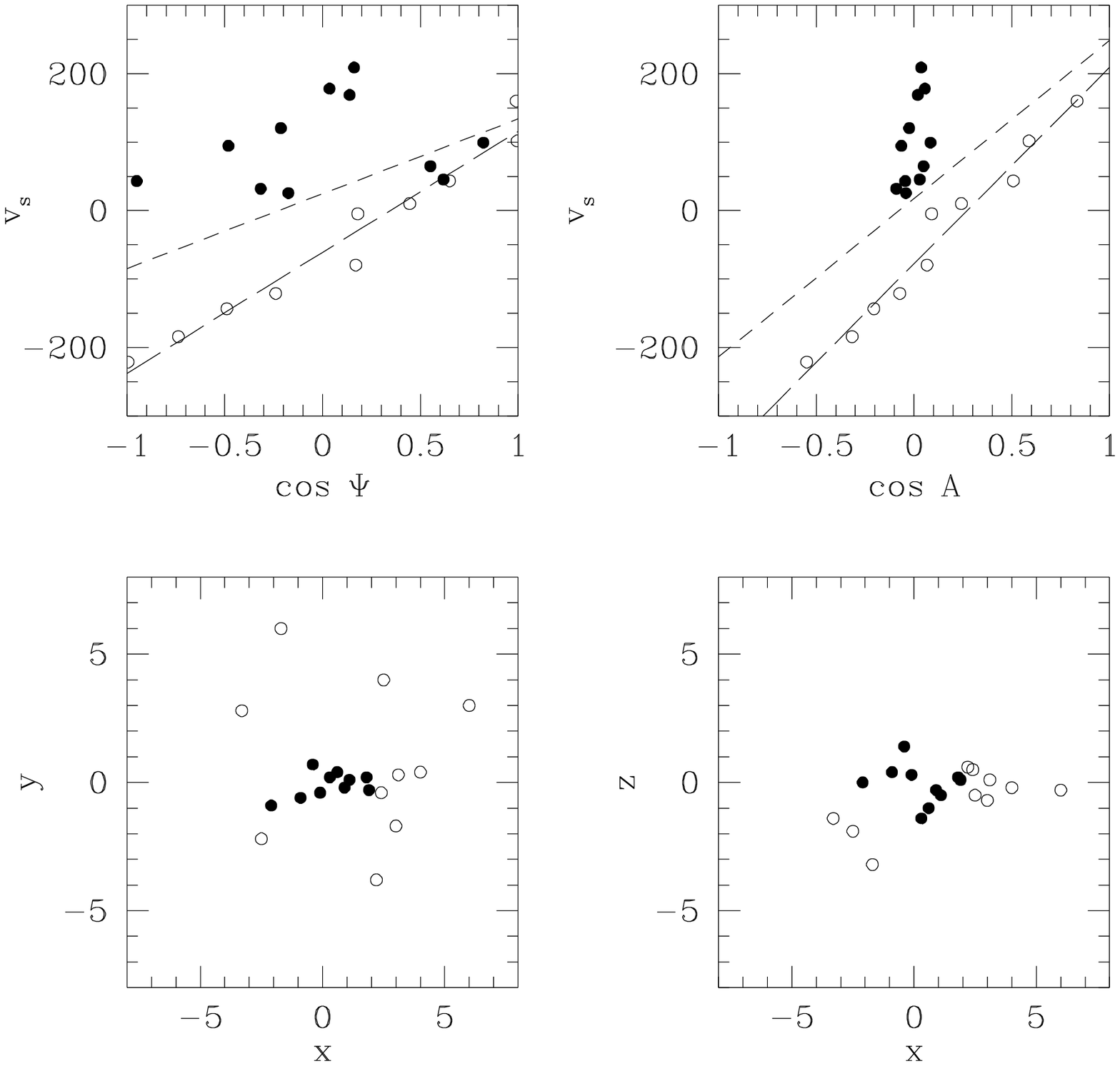]{The kinematical and spatial properties 
(upper and lower panels respectively) of Galactic 
globular clusters with metallicities of [Fe/H] $> -0.8$ and masses 
$\log (M/M_{\sun}) < 5.55$. The filled circles represent clusters of the 
``bar system'', while open circles denote the ``5-kpc-ring'' clusters.
The short-dashed lines show the least-squares fits to the velocity data for the
entire sample. The long-dashed lines show fits to the data for the 
``5-kpc-ring'' clusters only.} 

\figcaption[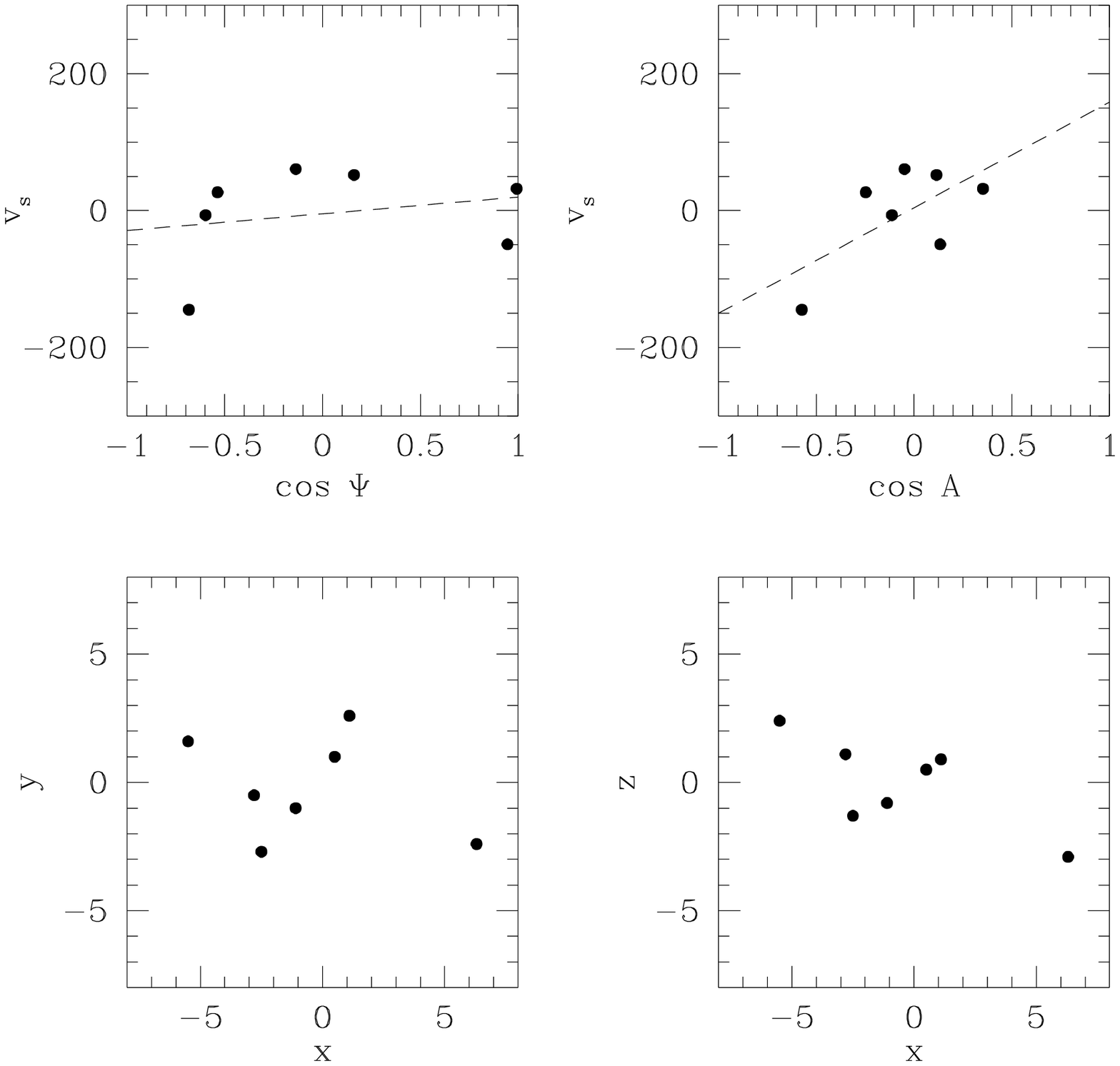]{The kinematical and spatial properties of Galactic 
globular clusters with metallicities of [Fe/H] $> -0.8$ and masses 
$\log (M/M_{\sun}) > 5.55$.}

\begin{deluxetable}{lcccc}
\scriptsize
\tablewidth{0pt}
\tablecaption{Kinematical and Spatial Properties of Globular Clusters 
    with [Fe/H] $> -0.8$}
\tablehead{     
\colhead{ }&\colhead{higher-mass}&\colhead{lower-mass}&\colhead{bar}& 
\colhead{5-kpc-ring}} 
\startdata
N & 7 & 21 & 11 & 10 \nl
$v_{\rm rot}$ [km/s] & $24.3 \pm 23.2$ & $113.2 \pm 14.0$ & $30.2 \pm 23.0$ &
                      $163.7 \pm 5.8$ \nl
$\omega$ [km/s/kpc] & $15.2 \pm 9.9$ & $24.3 \pm 5.3$ & $58.6 \pm 173.5$ &
                      $23.7 \pm 1.3$ \nl
$\sigma_{\rm los}$ [km/s] (flat) & 70.4 & 101.2 & 62.7 & 26.0 \nl
$\sigma_{\rm los}$ [km/s] (rigid) & 55.7 & 98.6 & 55.3 & 22.8 \nl
$Z_{\rm mean}$ [kpc] & 0.45 & 0.22 & 0.33 & 0.31 \nl
$R_{\rm mean}$ [kpc] & 3.49 & 3.18 & 2.18 & 4.27 \nl
\enddata
\end{deluxetable}

\end{document}